\documentclass[prl,twocolumn]{revtex4}
\usepackage{graphicx}
\usepackage{amssymb,amsmath,bm}
\usepackage{enumerate}

\newcommand{\be}{\begin{eqnarray}}
\newcommand{\ee}{\end{eqnarray}}

\begin{document}

\title{Superfluid (quantum) turbulence and distributed chaos}

\author{A. Bershadskii}

\affiliation{
ICAR, P.O. Box 31155, Jerusalem 91000, Israel
}

\begin{abstract}

Properties of distributed chaos in superfluid (quantum) turbulence have been studied using the data of recent direct numerical simulations (HVBK two-fluid model for He II, and a moving grid in the frames of Gross-Pitaevskii model of the Bose-Einstein condensates at low temperatures).
It is found that for the viscous (normal) component of the velocity field in He II the viscosity dominates the distributed chaos with the stretched exponential spectrum $\exp(-k/k_{\beta})^{\beta}$ and  $\beta = 2/3$. For the superfluid component the distributed chaos is dominated by the vorticity correlation integral with $\beta =1/2$ (the soft spontaneous breaking of the space translational symmetry - homogeneity). For very low temperature the distributed chaos is tuned to the large-scale coherent motions: the viscous (normal) component is tuned to the fundamental mode, whereas the superfluid component is subharmonically tuned. For the Gross-Pitaevskii superfluid turbulence incompressible part of the energy spectrum (containing the vortices contribution) also indicates the distributed chaos dominated by the vorticity correlation integral ($\beta =1/2$) and the subharmonic  tuning to the large-scale coherent motions. 
\end{abstract}

\maketitle

\section{Introduction}

In recent years high quality experimental and numerical simulations data became available for superfluid turbulence due to considerable progress in experimental and numerical simulations technique (see, for instance, Refs. \cite{nsd}-\cite{k} and references therein). These data provides a possibility  for an advance in our understanding of the superfluid turbulence. In particular, in a recent Ref. \cite{b1} it was shown, using the experimental data reported in the Ref. \cite{sal}, that distributed chaos plays a significant role in superfluid (He II) turbulent flow behind a grid. Therefore, it seems reasonable to investigate role of the distributed chaos in superfluid turbulence using also the data of the direct numerical simulations (DNS). Unlike the experimental data the data obtained in the DNS allow to separate between the viscous (normal) and inviscid (superfluid) components of velocity field (in the two-fluid model \cite{sclr},\cite{srl}) and between incompressible,
compressible and quantum components of the velocity field (for the Gross-Pitaevskii model of the Bose-Einstein condensates at low temperatures \cite{k}). This separation provides an additional information about underlying physical processes. \\

  The turbulence in the superfluid flows is as a rule inhomogeneous. Let us recall that for the distributed chaos with (a soft) spontaneous translational symmetry (homogeneity) breaking  \cite{b1} the vorticity correlation integral
$$
\gamma = \int_{V} \langle {\boldsymbol \omega} ({\bf x},t) \cdot  {\boldsymbol \omega} ({\bf x} + {\bf r},t) \rangle_{V}  d{\bf r} \eqno{(1)}
$$
dominates scaling of the group velocity of the waves driving the chaos
$$
\upsilon (\kappa ) \propto |\gamma|^{1/2}~\kappa^{\alpha} \eqno{(2)}
$$
with $\alpha = 1/2$ (from the dimensional considerations). Then the stretched exponential spectra of the distributed chaos  
$$
E(k )  \propto \exp-(k/k_{\beta})^{\beta}  \eqno{(3)}
$$
have 
$$
\beta =\frac{2\alpha}{1+2\alpha} = 1/2  \eqno{(4)}
$$ 
in this case.

\section{Two-fluid model for the superfluid He II}

A liquid $~^4$He goes through a phase transition at $T_{\lambda} \simeq 2.17$ K and at 
$T < T_{\lambda}$ it becomes a superfluid He II. The two-fluid model was suggested by 
L. Landau \cite{lan} for a phenomenological description of the superfluid He II. It is suggested in this model that the He II can be seen as a superposition of a viscous (normal - with the Navier-Stokes dynamics) and supeflud (Euler dynamics with quantized vorticity) components coupling by a friction term. Then this model was developed into so-called truncated Hall-Vinen-Bekeravich-Khalatnikov (HVBK \cite{vn}) model with a cut-off associated with the quantum scale $\delta$ (an inter-vortex spacing) \cite{srl}. The most of information about the {\it quantized} vorticity (such as the Kelvin waves, for instance) is smoothed out by the scale $\delta$ in this truncated model. Therefore, the main physics is concentrated here in the mutual coupling term for the Navier-Stokes and Euler equations \cite{srl}. 

  Before considering results of a DNS based on this model let us specify power spectrum expected for distributed chaos corresponding to the viscous (normal) component of the velocity field. For the viscous component just viscosity - $\nu $, is the parameter that should determine the scaling Eq. (2) instead of the parameter $\gamma$. Then from the dimensional consideration we obtain
$$
\upsilon (\kappa ) \propto \nu~\kappa^{\alpha} \eqno{(5)}
$$
with $\alpha =1$. Substituting this value of $\alpha$ into Eq. (4) one obtains
$$
\beta =\frac{2\alpha}{1+2\alpha} = 2/3  \eqno{(6)}
$$ 
  
  In Refs. \cite{sclr},\cite{srl} results of direct numerical simulations on the basis of the above described model (in cubic box with periodic boundary conditions, resolution $512^3$ for T = 2.1565 K and $1024^3$ for T = 1.15 K) were reported. Stationarity of the turbulence is provided by an external force (isotropic) at scale  $L_0$. 
  
  Figure 1 shows the power spectrum obtained in these simulations (the data were taken from Fig. 5 in the Ref. \cite{sclr}) for the viscous (normal) component of the velocity field at T=2.1565 K. The dashed straight line indicates (in the scales chosen for this figure) the stretched exponential spectrum Eq. (3) with the $\beta =1/3$ given by Eq. (6). 
  \begin{figure}
\begin{center}
\includegraphics[width=8cm \vspace{-1.17cm}]{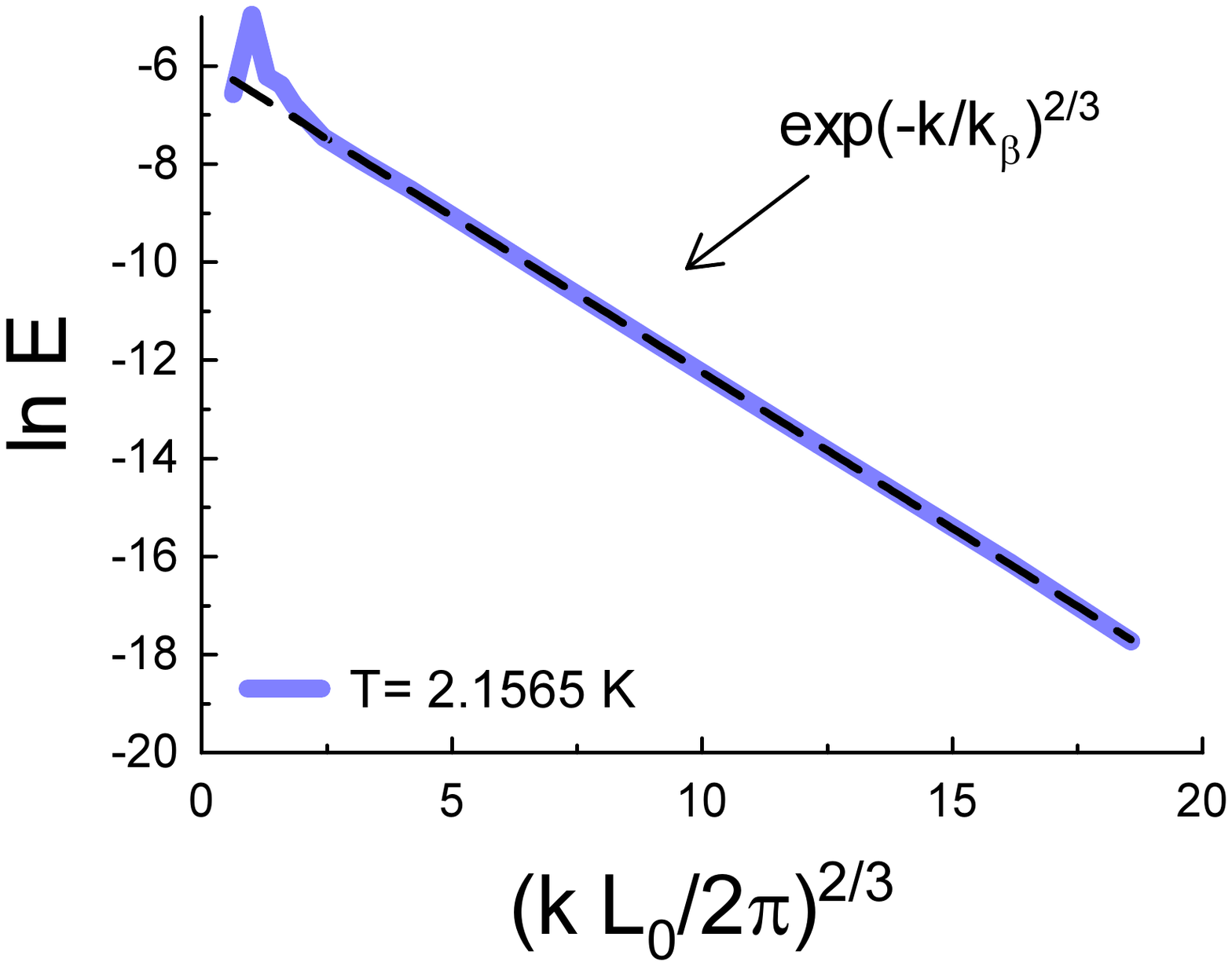}\vspace{-3.2cm}
\caption{\label{fig1}  The power spectrum for the viscous (normal) component of the velocity field at T=2.1565 K (the data were taken from Fig. 5 in the Ref. \cite{sclr}).  }
\end{center}
\end{figure}
\begin{figure}
\begin{center}
\includegraphics[width=8cm \vspace{-1.9cm}]{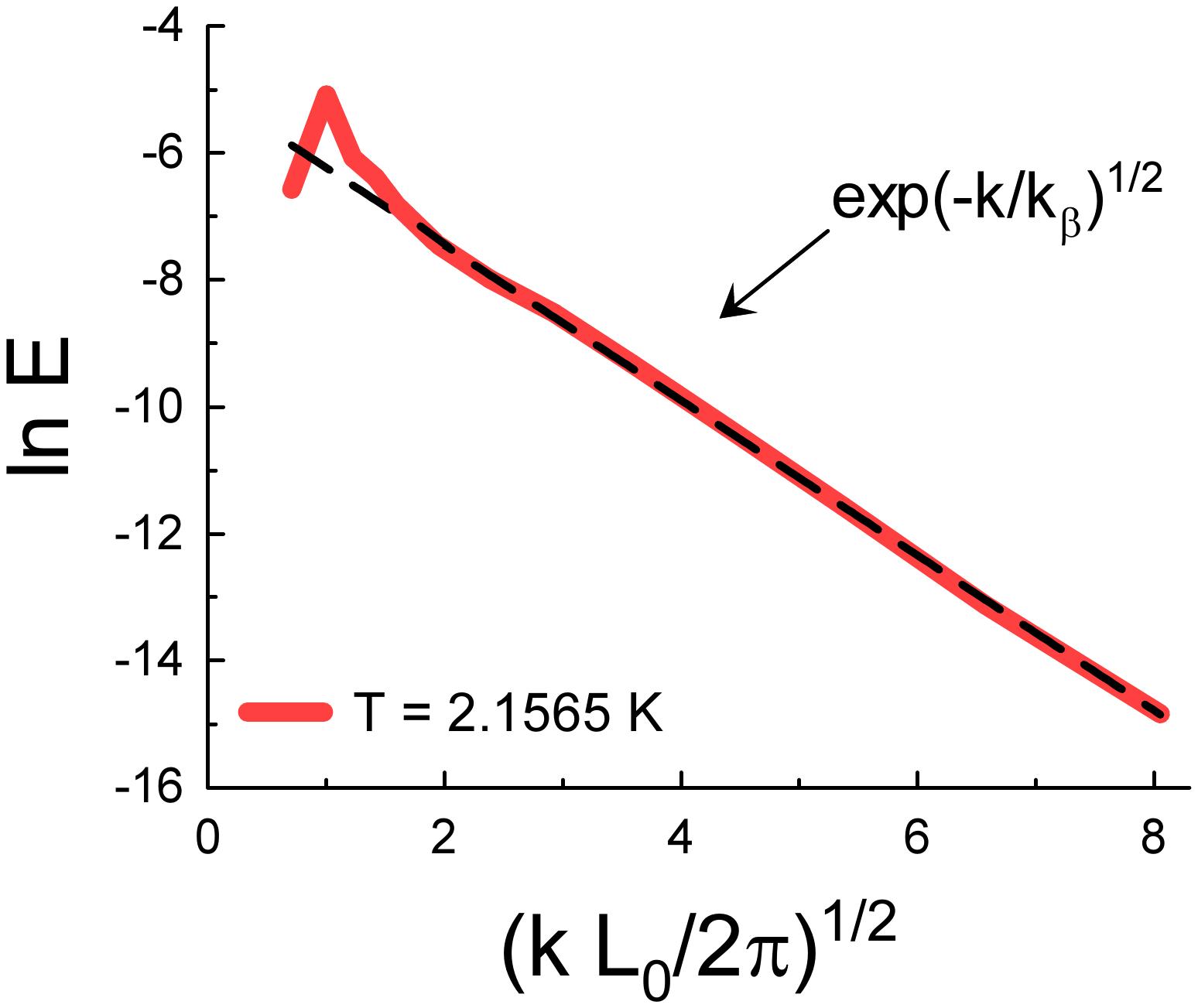}\vspace{-2.3cm}
\caption{\label{fig2} The same as in Fig. 1 but for the data obtained for the superfluid component of the velocity field. The $\beta$ is changed from 2/3 to 1/2.} 
\end{center}
\end{figure}
\begin{figure}
\begin{center}
\includegraphics[width=8cm \vspace{-1.1cm}]{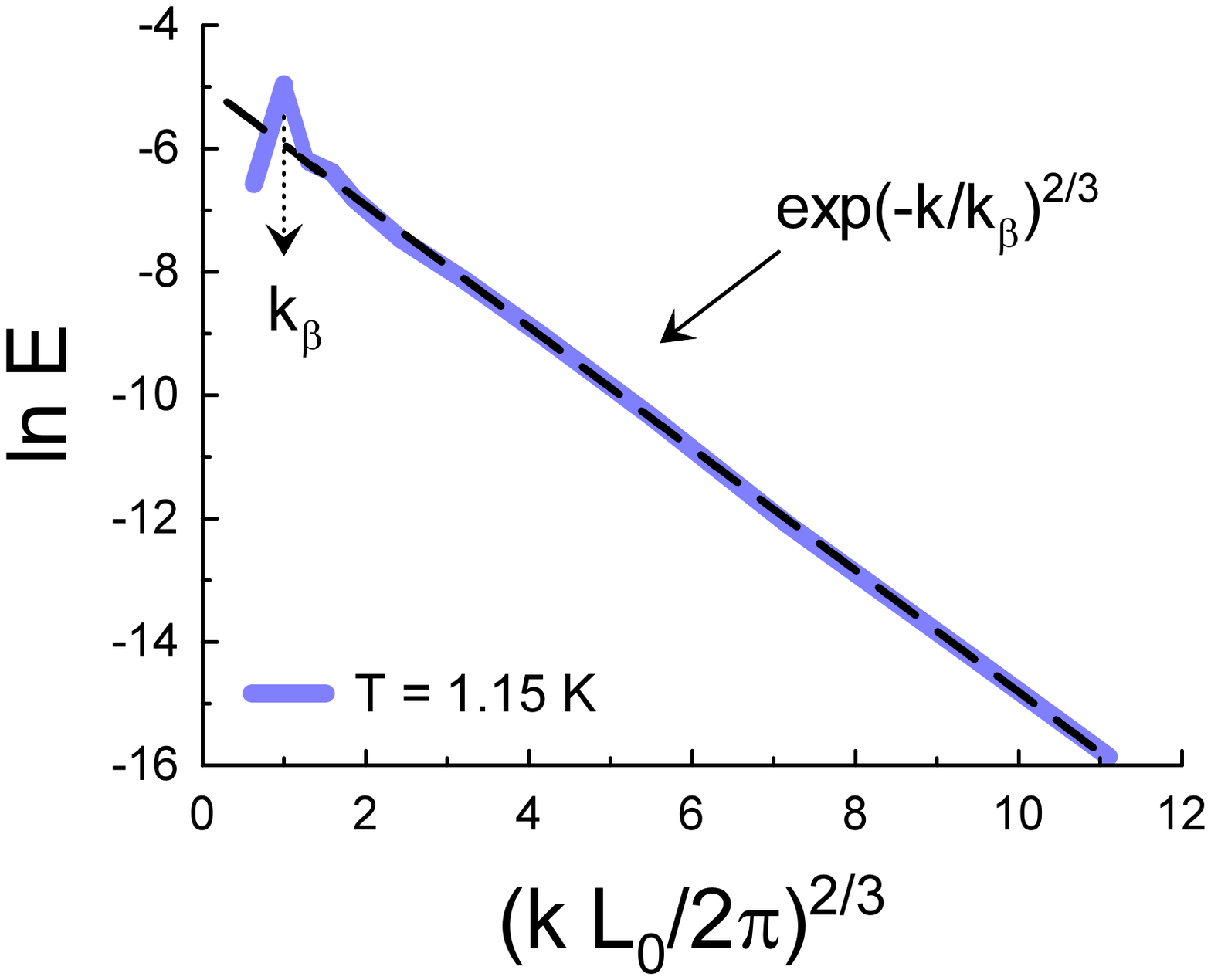}\vspace{-3.3cm}
\caption{\label{fig3} The same as in Fig. 1 but for lower T = 1.15 K. The arrow under the peak indicates tuning of the $k_{\beta}$ to the wavenumber equals to position of the spectral peak. }
\end{center}
\end{figure}
  Figure 2 shows analogous data but obtained for the inviscid (superfluid) component of the velocity field. The dashed straight line indicates (in the scales chosen for this figure) the stretched exponential spectrum Eq. (3) with the $\beta =1/2$ given by Eq. (4). One can see that unlike the case of the viscous (normal) component the distributed chaos for the superfluid component is determined by the soft spontaneous breaking of the space translational symmetry (homogeneity), with domination of the vorticity correlation integral Eq. (1). One should remember that the viscosity have an influence on the superfluid component of the velocity field due to the mutual coupling terms (friction).

\section{Tuning to the large scale coherent motions}

  Figure 3 shows the power spectrum for viscous (normal) component of velocity field obtained in the same DNS as that shown in Fig. 1 but for very low temperature T = 1.15 K. As in Fig. 1 
the dashed straight line indicates the stretched exponential spectrum Eq. (3) with the $\beta =1/3$ given by Eq. (6). However, there is something new in this case. The stretched exponential part of the spectrum with $\beta =2/3$ exhibits tuning of the $k_{\beta}$ to the wavenumber equal to position of the spectral peak (see the arrow under the peak in Fig. 3). 

   The tuning in the case of the superfluid component is even more interesting at this low temperature. In this case the distributed chaos inherits a property of deterministic chaos - subharmonic tuning. Let us consider the famous Lorenz equations
$$
\frac{dx}{dt} = \sigma (y - x),~~      
\frac{dy}{dt} = r x - y - x z, ~~
\frac{dz}{dt} = x y - b z      \eqno{(7)}          
$$
It is an oversimplified model for classic thermal convection \cite{lorenz}. Figure 4 shows power spectrum of $z$-component of the Lorenz strange attractor (deterministic chaos at $\sigma=10.0,~ r = 28.0,~ b = 8/3$). The maximum entropy method was used for computing this spectrum.  The straight line indicates exponential spectrum typical for the smooth deterministic chaos. The tuning in this case is subharmonic (period-doubling): the fundamental frequency (the first spectral peak) equals to $2f_{\beta}$ (see arrow under the first peak). 

  Figure 5 shows the power spectrum for superfluid component of velocity field obtained in the same DNS as that shown in Fig. 2 but for the lower temperature T = 1.15 K. As in Fig. 2 
the dashed straight line indicates the stretched exponential spectrum Eq. (3) with the $\beta =1/2$ given by Eq. (4) (domination of the vorticity correlation integral Eq. (1)). The arrow under the peak indicates the subharmonic tuning of the distributed chaos.  

  The tuning observed at the low temperature T = 1.15 K can be related to increase in coherence for lower temperatures. 
  
\section{Gross-Pitaevskii turbulence generated by a grid}   
\begin{figure}
\begin{center}
\includegraphics[width=8cm \vspace{-1cm}]{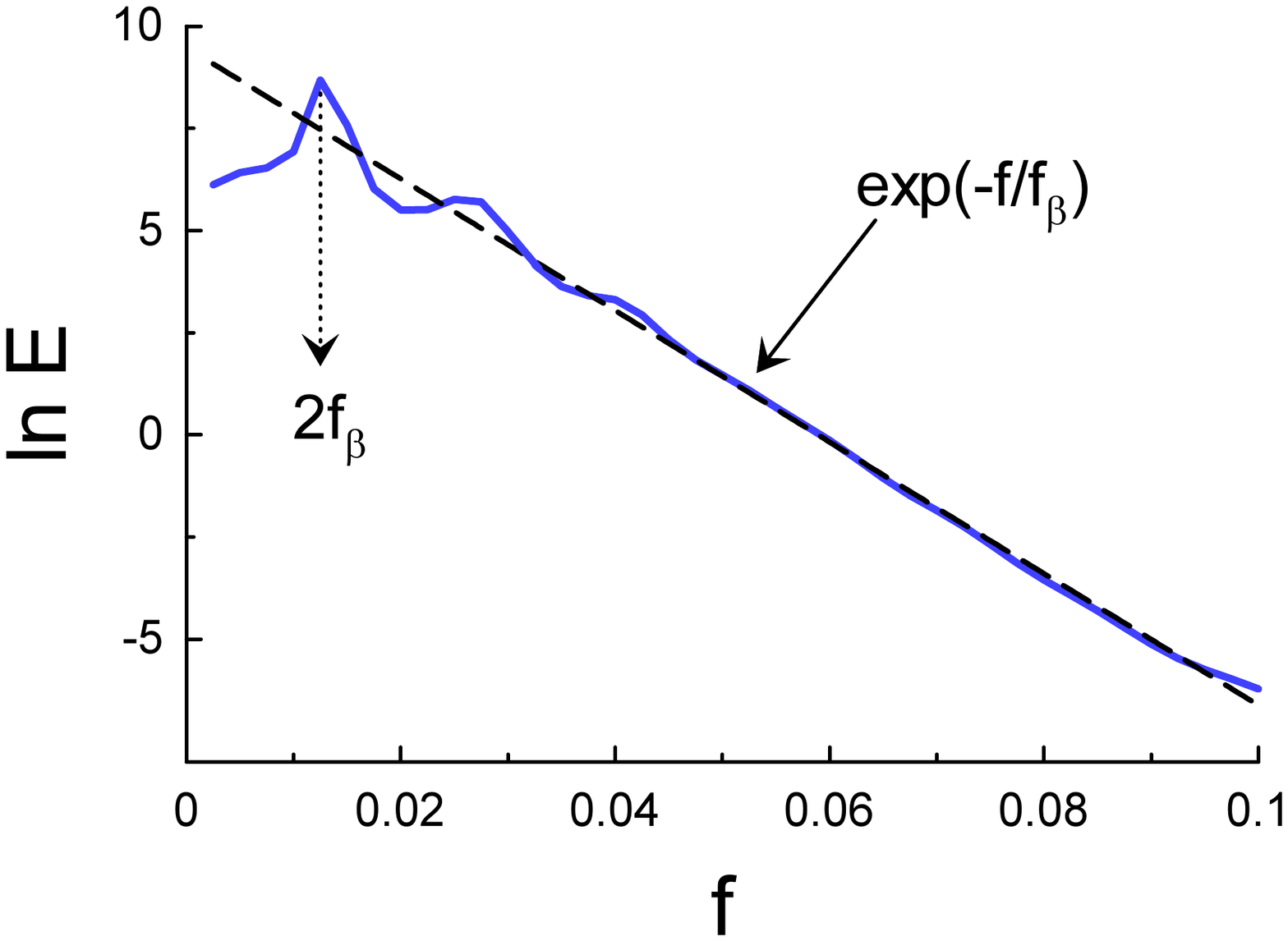}\vspace{-4.3cm}
\caption{\label{fig4}  The power spectrum of $z$-component of the Lorenz strange attractor. The arrow under the first peak indicates the subharmonic tuning. } 
\end{center}
\end{figure}
\begin{figure}
\begin{center}
\includegraphics[width=8cm \vspace{-1.75cm}]{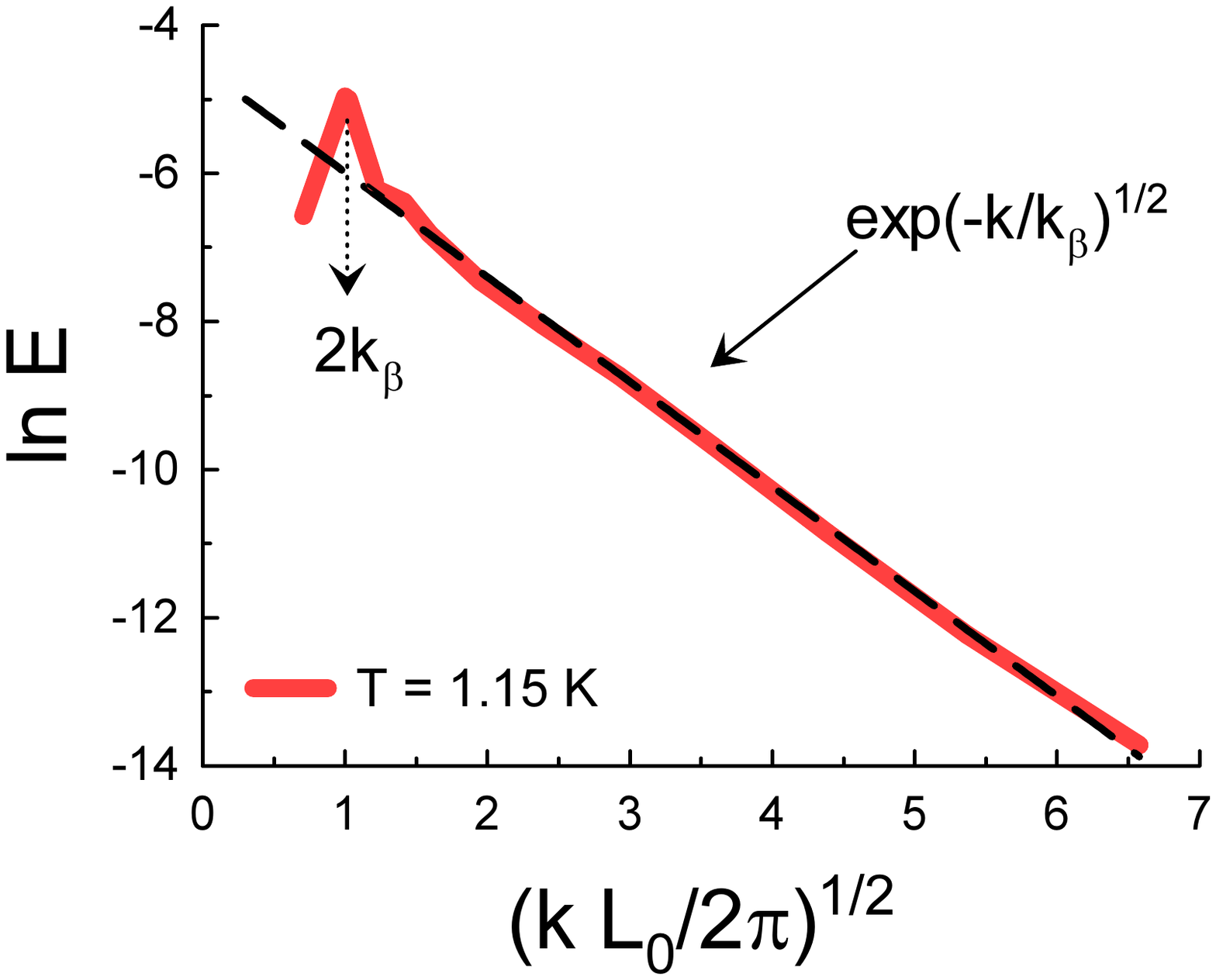}\vspace{-2.8cm}
\caption{\label{fig5} The same as in Fig. 2 but for lower T = 1.15 K. The arrow under the peak indicates the subharmonic tuning. }
\end{center}
\end{figure}
\begin{figure}
\begin{center}
\includegraphics[width=8cm \vspace{-1.1cm}]{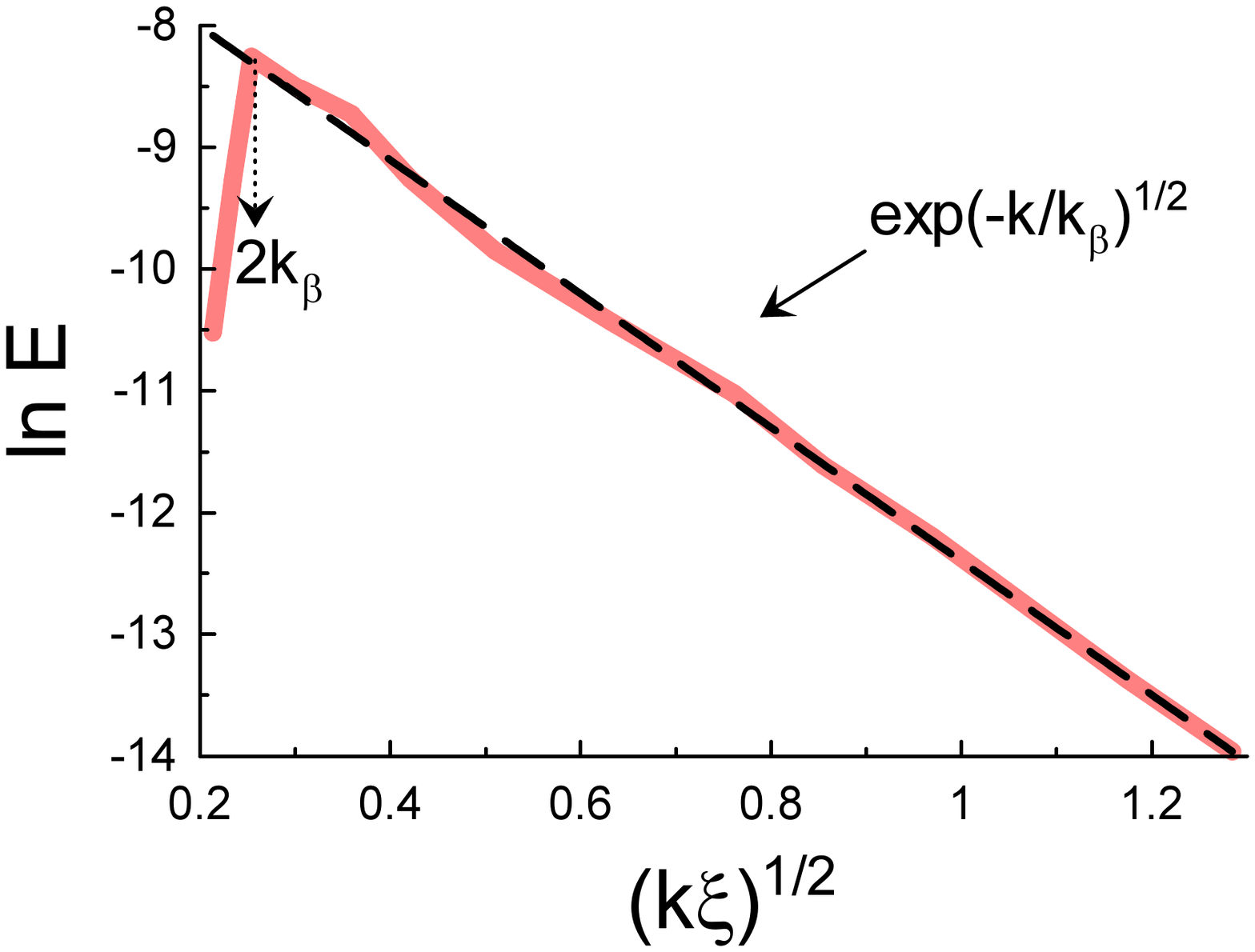}\vspace{-4.2cm}
\caption{\label{fig6} The incompressible part of the energy spectrum. The data were taken from Fig. 2c of the Ref. \cite{k}. The arrow under the peak indicates the subharmonic tuning. }
\end{center}
\end{figure}
  A very interesting DNS was reported in a recent Ref. \cite{k}. In this DNS superfluid turbulence was generated by a moving grid in the frames of the Gross-Pitaevskii equation
$$
i\hbar \frac{\partial \psi}{\partial t} =- \frac{\hbar^2}{2m} \nabla^2 \psi + g|\psi|^2\psi  \eqno{(8)}
$$ 
with $m$ as the condensed particles mass. This is a non-linear Schr\"{o}dinger equation for wave-function $\psi$ usually used for the low temperature limit description of the Bose-Einstein condensates. The superfluid velocity is obtained by the Madelung's transformation $\psi({\bf x},t)=\sqrt{\frac{\rho({\bf x},t)}{m}}\exp{[i \frac{m}{\hbar}\phi({\bf x},t)]}$, with $\rho({\bf x},t)$ as density of the superfluid and ${\bf v} = \nabla \phi$ as its velocity field (the domain generally is not simply connected, see also \cite{b2} for a generalization of vorticity and corresponding adiabatic invariant).

  In this DNS \cite{k} a repulsive potential $V_{gr}({\bf x})$ simulates the grid and the mean flow is simulated by addition of an advection term $\vec{v_0}\cdot\nabla\psi$ into the Gross-Pitaevskii equation.
  
   The author of the Ref. \cite{k} reports about two stage development of the wake behind the grid. The first stage is characterized by injection of incompressible kinetic energy due to 
nucleation of the vortex rings generated by the moving grid. At certain effective Mach number the vortex ring nucleation saturates and stops. Before this saturation an inhomogeneous turbulent state is observed in the DNS. In the second stage the  mutual friction results in the rings
shrinking.

  Figure 6 shows the incompressible part of the turbulent energy spectrum - containing
the vortices contribution (distance from the grid $\simeq 360 \xi$, where $\xi$ is the coherence
length, which is approximately equal to the vortex core size). As in Fig. 5
the dashed straight line indicates the stretched exponential spectrum Eq. (3) with the $\beta =1/2$ given by Eq. (4) (domination of the vorticity correlation integral Eq. (1)). The arrow under the peak indicates the subharmonic tuning of the distributed chaos.  
  
  While the above used HVBK model smoothed out the quantized
vortices the Gross-Pitaevskii model provides more details for description of the quantum turbulence. Although the Gross-Pitaevskii model was primarily intended for the Bose-Einstein condensates at low temperatures, one can expect that it can be useful 
for superfluid turbulence in Helium as well (cf. Figs. 5 and 6).\\

  I thank K. R. Sreenivasan for introducing me into the challenging problems of superfluid turbulence.

\end{document}